\documentclass[manuscript]{biometrika}

\usepackage{setspace}

\usepackage{amsmath}
\usepackage{amssymb}
\usepackage{times}
\usepackage{bm}
\usepackage{color}
\usepackage{multicol}
\usepackage{multirow}
\usepackage{mathrsfs}  
\usepackage[labelsep=period]{caption}
\usepackage{subcaption}
\usepackage{booktabs}
\usepackage[flushleft]{threeparttable}
\usepackage{wasysym}
\usepackage{pdflscape}
\usepackage{enumitem}
\usepackage[plain,noend]{algorithm2e}
\usepackage{setspace}
\usepackage{graphicx}
\usepackage{listings}

\lstnewenvironment{rc}[1][]{\lstset{language=R}}{}

\makeatletter
\renewcommand{\algocf@captiontext}[2]{#1\algocf@typo. \AlCapFnt{}#2} 
\def\@algocf@capt@plain{top}
\renewcommand{\algocf@makecaption}[2]{%
  \addtolength{\hsize}{\algomargin}%
  \sbox\@tempboxa{\algocf@captiontext{#1}{#2}}%
  \ifdim\wd\@tempboxa >\hsize
    \hskip .5\algomargin%
    \parbox[t]{\hsize}{\algocf@captiontext{#1}{#2}}
  \else%
    \global\@minipagefalse%
    \hbox to\hsize{\box\@tempboxa}
  \fi%
  \addtolength{\hsize}{-\algomargin}%
}
\makeatother

\begin{document}

\newcommand\independent{\protect\mathpalette{\protect\independenT}{\perp}}
\def\independenT#1#2{\mathrel{\rlap{$#1#2$}\mkern2mu{#1#2}}}

\newcommand{\CSPS}{{\textsc{csps}}}
\renewcommand{\thesubfigure}{\roman{subfigure}}
\newcommand{\beq}[1]{\begin{equation}\label{#1}}
\newcommand{\eeq}{\end{equation}}
\newcommand{\abs}[1]{\left|#1\right|}
\newcommand{\paren}[1]{(#1)}
\newcommand{\sqbrkt}[1]{\left[#1\right]}
\newcommand{\set}[1]{\left\{#1\right\}}
\newcommand{\mb}[1]{\boldsymbol{#1}}
\newcommand{\mbt}[1]{\mb{#1}}
\newcommand{\E}[1]{E(#1)}
\newcommand{\Prob}[1]{\mathop{}\!\mathrm{pr}(#1)}

\newcommand{\clam}[1]{c (#1)}
\newcommand{\clamp}[1]{c_{\lambda'}(#1)}
\newcommand{\clami}[1]{c_{j}(#1)}
\newcommand{\Dlam}{D_{\lambda}}
\newcommand{\Dlamn}{D_{n\lambda}}
\newcommand{\Dlamp}{D_{\lambda'}}
\newcommand{\tmx}{\mb{X}}
\newcommand{\tmw}{\mb{W}}
\newcommand{\tmy}{\mb{Y}}
\newcommand{\tx}{X}
\newcommand{\ty}{Y}
\newcommand{\tw}{W}
\newcommand{\tmdl}{\mb{D}}
\newcommand{\tdl}{D}
\newcommand{\tdln}{D_{n}}
\newcommand{\tmd}{\mb{D}}
\newcommand{\td}{D}
\newcommand{\ind}[1]{\boldsymbol{1}_{#1}}

\newcommand{\enumT}{t =1, \cdots, T}
\newcommand{\enumN}{n =1, \cdots, N}
\newcommand{\bcol}[1]{{\color{blue} #1}}
\newcommand{\rcol}[1]{{\color{red} #1}}

\markboth{S. Han and D. B. Rubin}{}

\title{Contrast Specific Propensity Scores}

\author{S. HAN}
\affil{Beijing International Center for Mathematical Research, Peking University \& Department of Health Care Policy, Harvard University, No. 5 Yiheyuan Road Haidian District, Beijing 100871, China.\email{shashahan@pku.edu.cn}}

\author{D. B. RUBIN}
\affil{Yau Mathematical Center, Tsinghua University \& Department of Statistical Science, Temple University, Jin Chun Yuan West Building, Haidian District, Beijing 100084, China.
\email{rubin@stat.harvard.edu}}

\maketitle

\begin{abstract}
Basic propensity score methodology is designed to balance multivariate pre-treatment covariates when comparing one active treatment with one control treatment. Practical settings often involve comparing more than two treatments, where more complicated contrasts than the basic treatment-control one, $(1,-1)$, are relevant.  Here, we propose the use of contrast-specific propensity scores (\CSPS{}). \CSPS{} allow the creation of treatment groups of units that are balanced with respect to bifurcations of the specified contrasts and the multivariate space spanned by them.

\end{abstract}
\begin{keywords}
Causal inference; Covariate balance; Multiple treatments.
\end{keywords}
\section{Introduction}

Studies with multiple treatments, whether due to multi-valued treatments (e.g., doses of a drug) or many factors (e.g., several types of drugs), are common. A \emph{contrast} among $T$ treatments is a vector of $T$ coefficients that sum to zero. With two treatments, there is only one contrast, conventionally written as $(1, -1)$, but with more than two treatments, there are multiple contrasts. Contrasts have long been used in experimental studies (see \citealp{SC67, RRR00,WH11}). Generally, a contrast-specific propensity score results from a bifurcation, that is, a partition of treatments into two groups. 

When $T=2$, \cite{RR83} proposed the use of the propensity score to balance multivariate covariates, and it is now widely used. The propensity score is the conditional probability of assignment to treatment versus control, given pre-treatment covariates. Some extensions of the propensity score to studies with multiple treatments have been suggested. The ``generalized propensity score'' proposed by \cite{Imbens00} is primarily applicable to methods using inverse probability weighting estimation \citep{HT52}. The multidimensional propensity score proposed by \cite{Lechner01} and the propensity function proposed by \cite{ID04} differ from what we propose.

A contrast-specific propensity score is the conditional probability of assignment to a bifurcation of treatment groups, for example, the treatment groups having positive coefficients versus having negative ones in that contrast, given pre-treatment covariates. We propose the use of contrast-specific propensity scores (\CSPS{}) to create treatment groups with balanced covariate distributions in the multidimensional space spanned by bifurcations of contrasts.

\section{Contrasts among treatments}\label{sec: notation:contrasts}

A $T$ component vector $\mb{\lambda} =(\lambda_1, \ldots, \lambda_T)$ is a contrast if $\sum_{t= 1}^T \lambda_t = 0$. The sign of contrast $\mb{\lambda}$ is $ \text{sgn}(\mb{\lambda}) = (\text{sgn}(\lambda_1), \ldots, \text{sgn}(\lambda_T))$, with $\text{sgn}(\lambda_t) = \lambda_t/\abs{\lambda_t} $ if $\lambda_t \neq 0$ and $\text{sgn}(\lambda_t) = 0$ if $\lambda_t = 0$. Two contrasts $\mb{\lambda}_1$ and $\mb{\lambda}_2$ are orthogonal if $\sum_{t=1}^T \lambda_{1t}\cdot \lambda_{2t} = 0$. 

Let $\mb{\lambda}^+ =(\lambda_1^+, \ldots, \lambda_T^+)$ and $\mb{\lambda}^-=(\lambda_1^-, \ldots, \lambda_T^-)$ denote the nonnegative and nonpositive components of $\mb{\lambda}$ respectively, i.e., $\lambda_t^+ = \max\set{\lambda_t, 0}$ and $\lambda_t^- = \min\set{\lambda_t, 0}$. 
The sgn function bifurcates contrast $\mb{\lambda}$ into $\set{\text{sgn}(\mb{\lambda}^+), \text{sgn}(\mb{\lambda}^-)}$. For example, it bifurcates contrast $(1/2,1/2,-1)$ into $\set{(1, 1, 0), (0, 0, -1)}$. 

The $ \text{sgn}$ function bifurcates a contrast using the zero-boundary for all $T$ components. Non-zero boundaries can be appropriate, depending on the goal of investigation. Generally, let $f(\mb{\lambda}) = (f(\lambda_1), \ldots, f(\lambda_T))$ be the function for bifurcation, with $ f(\lambda_t)=  \lambda_t/\abs{\lambda_t} $ if $\lambda_t < \ell_t $ or $ \lambda_t >  u_t$ and $f(\lambda_t) = 0$ otherwise, where $\ell_t$ and $u_t$ are the lower and upper boundaries for component $t$. Bifurcation function $f$ bifurcates contrast $\mb{\lambda}$ into $\set{f(\paren{\mb{\lambda}}^+), f(\paren{\mb{\lambda}}^-)}$. For example, for the linear contrast with four groups, $(-3, -1, +1, +3)$, the bifurcation function $f$ with the lower boundary $(-1, -1, -1, -1)$ and upper boundary $(1, 1, 1, 1)$ bifurcates the contrast into $\set{(-1, 0, 0, 0), (0, 0, 0, 1)}$.

\begin{example}\textbf{(One active treatment with two control conditions)}\label{eg many contrl}
For one active treatment with two controls, the contrast $(1, -1/2, -1/2)$ compares the active treatment to the average of the two controls, and the contrast $(0, 1, -1)$ compares the two controls. Contrasts $(1, -1/2, -1/2)$ and $(0, 1, -1)$ are orthogonal. For instance, \cite{Lalonde86} was interested in the contrast of one experimental treatment, a job training programme, and the average of two controls, using the non-experimental groups from the Panel Study of Income Dynamics and from the Current Population Survey. 
\end{example}

\begin{example}\textbf{(Multiple factors each with two levels)}\label{eg many factors}
Table \ref{tab: many factor} displays the case of three factorial treatments, each with two levels denoted by $0$ and $1$. The contrasts $\mb{\lambda}_j$, $j = 1, 2, 3$, define the three main effects, the contrasts $\mb{\lambda}_j$, $j = 4, 5, 6$, define the three two-way interaction effects, and the contrast $\mb{\lambda}_7$ defines the three-way interaction effect \cite[see, e.g.,][]{SC67}. The contrast $\mb{\lambda}_8$ compares the combination of factors A and B versus the main effect of factor A; the contrast $\mb{\lambda}_9$ compares the combination of factors A and B versus the main effect of factor B; the contrast $\mb{\lambda}_{10}$ compares the effect of factors A and B both at level ``1" versus when they are both at level ``0". 
For example, in a recent study, \cite{KMW+19} considered the contrasts $(-1, 1, 0, 0)$ and $(0, 0, -1, 1)$.

\end{example}
\begin{table}[!htp]
\small
\centering
\caption{Three factorial treatments, each with two levels}\label{tab: many factor}
\begin{tabular}{cccccccccc}
\hline
Treatments Indexings & & 1  & 2  & 3  & 4  & 5  & 6  & 7  & 8 \\ \hline
\multirow{3}{*}{Factors} & A  & 0  & 0  & 0  & 0  & 1  & 1  & 1  & 1 \\ \cline{3-10}
                                   & B   & 0  & 1  & 0  & 1  & 0  & 1  & 0  & 1 \\ \cline{3-10}
                                   & C    & 0  & 0  & 1  & 1  & 0  & 0  & 1  & 1 \\ \hline
\multirow{7}{*}{Contrasts} &$\mb{\lambda}_1$       & -1 & -1 & -1 & -1 & ~1  & ~1  & ~1  & 1 \\ \cline{3-10}
                    &$\mb{\lambda}_2$       & -1 & ~1  & -1 & ~1  & -1 & ~1  & -1 & 1 \\ \cline{3-10}
                    &$\mb{\lambda}_3$       & -1 & -1 & ~1  & ~1  & -1 & -1 & ~1  & 1 \\ \cline{3-10}
                    &$\mb{\lambda}_4$       & ~1  & -1 & ~1  & -1 & -1 & ~1  & -1 & 1 \\ \cline{3-10}
                    &$\mb{\lambda}_5$       & ~1  & -1  & -1 & ~1  & ~1  & -1 & -1 & 1 \\ \cline{3-10}
                    &$\mb{\lambda}_6$       & ~1  & ~1  & -1 & -1 & -1 & -1 & ~1  & 1 \\\cline{3-10}
                    &$\mb{\lambda}_7$       & -1 & ~1  & ~1  & -1 & ~1  & -1 & -1 & 1 \\ \cline{3-10}
                    &$\mb{\lambda}_8$       & ~0 & ~0  & ~0  & ~0 & -1  & ~1 & -1 & 1 \\ \cline{3-10}
                    &$\mb{\lambda}_9$       & ~0 & -1  & ~0  & -1 & ~0  & ~1 & ~0 & 1 \\  \cline{3-10}
                   &$\mb{\lambda}_{10}$       & -1 & ~0  & -1  & ~0 & ~0  & ~1 & ~0 & 1 \\ \hline
\end{tabular}
\end{table}

\section{Contrast specific propensity scores using the $ \text{sgn}$ bifurcation}

\subsection{Basic notation}\label{sec: notation}
Consider a study with $N$ units, indexed by $i \in \set{ 1, \ldots, N}$.  The outcome variable $Y$ is measured on each unit after its treatment exposure. Associated with treatment $t$ is the potential outcome $Y_i(t)$, the value of $Y$ when the unit $i$ is exposed to treatment $t$, which implicitly assumes the stable unit treatment value assumption (SUTVA) \citep{Rubin80b}, within the ``Rubin Causal Model'' \citep{Rubin74} --- often called the ``potential outcomes approach to causal inference'' \citep{IR15}. The $\mb{\lambda}$ contrast of potential outcomes for unit $i$  is $ \sum_{t = 1}^T \lambda_t Y_i(t)$. Each unit $i$ is associated with covariates $X_i$, $X_i = (X_{i1}, \ldots,X_{iK} )\in \mathbb{R}^K$, that are measured prior to treatment exposure, and which ideally are balanced across treatment groups. 

\subsection{Contrast specific propensity scores using the $ \text{sgn}$ bifurcation}\label{sec: csps defi}

 \CSPS{} are designed to create treatment groups with balanced multivariate covariate distributions in the subspace spanned by bifurcations of contrasts among multiple treatments. Let $w_{it}$ be the indicator for whether unit $i$ is assigned to treatment $t$. Specifically,  $w_{it} = 1$ if $W_i = t$ and 0 otherwise, where $W_i = t$ indicates that unit $i$ receives treatment $t$, $t \in \set{1, \ldots, T}$. Let $\tdl_i$ be the indicator that whether $w_{it}$ corresponds to a positive, zero or negative $\lambda_t$, i.e., $\tdl_i = \sum_{t=1}^T \text{sgn}(\lambda_t) \cdot w_{it}$.  Based on $D_i$, \CSPS{} are the conditional probability of assignment to the treatment groups with positive coefficients versus negative coefficients of contrast $\mb{\lambda}$, given pre-treatment covariates $X_i$,
\beq{eq: csps}
\clam{X_i} = \Prob{D_i = 1 \mid X_i, \abs{D_i} = 1}.
\eeq

Let the conditional probability of assignment to treatment $t$, given covariates $X_i$, be $p_t(X_i) = \Prob{W_i = t |X_i}$. In Example \ref{eg many contrl}, \CSPS{} for the contrast $(1, -1/2, -1/2)$ is $p_1(X_i)$; In Example \ref{eg many factors}, \CSPS{} for the contrast $\mb{\lambda}_1$ is $\sum_{t=4}^8 p_t(X_i)$, which equals $\Prob{A = 1 \mid X_i}$.

\subsection{Unconfoundedness}\label{sec: assump}
The assignment with respect to the sgn bifurcation of contrast $\mb{\lambda}$ is unconfounded, given covariates $X_i$, if 
$
D_i  \independent \ty_i(1), \ldots,  \ty_i(T) \mid X_i. 
$
The condition is weaker than the strong unconfoundedness condition $W_i \independent \ty_i(1), \ldots,  \ty_i(T)  \mid X_i $, stated in \cite{IR15}. 
 
\subsection{Balancing Properties}\label{sec: theories}

As with the basic propensity score, the key advantage of \CSPS{}  is their balancing property. 
\begin{property}\label{theorem blance CSPS}
	Balance on $\clam{X_i}$ of the bifurcation of contrast $\mb{\lambda}$ balances the bifurcation of that contrast, i.e., $D_i  \independent X_i \mid \paren{\clam{X_i}, \abs{D_i} = 1} $, and therefore the subspace spanned by the bifurcation of contrast $\mb{\lambda}$. 
\end{property}

An immediate implication is that balancing on $p_t(X_i)$ balances the subspace spanned by sgn bifurcations of contrast $\lambda_t = 1, \lambda_{t'} =  -1/(T-1)$, $t' \neq t, t' = 1, \ldots, T$. For $J$ contrasts, let $\clami{X_i}$ be the \CSPS{} of the contrast $\mb{\lambda}_j$ and $D_{ij}$ the indicators with respect to contrast $\mb{\lambda}_j$, $j =1, \ldots, J$. We find,
\begin{property}\label{theorem blance L CSPS}
	Balance on $\clami{X_i}$, or any one-to-one function of $\clami{X_i}$, $j = 1, \ldots,J $, balances the bifurcations of contrasts $\mb{\lambda}_j, j = 1, \ldots,J $. That is, $ (\td_{ij}; j = 1, \ldots, J)  \independent X_i \mid (\clami{X_i}, \abs{D_{ij}} = 1; j = 1, \ldots,J)$, and therefore balances the subspace spanned by these bifurcations of contrasts $\mb{\lambda}_1, \ldots,\mb{\lambda}_J$. 
\end{property}

Because $T-1$ linearly independent vectors span the full space, we have that balance on \CSPS{} of contrasts with $T-1$ linearly independent sign vectors, e.g., sgn bifurcations of orthogonal contrasts, balances all linear contrasts among $T$ treatments, and all bifurcations among $T$ treatments. 

Property \ref{theorem blance L CSPS} has an interesting implication in practice. The $J$ \CSPS{}, once created, could be treated as covariates using a usual propensity score analysis, which creates a chain of balance. That is, balance on the propensity scores with the $J$ \CSPS{} used as covariates balances the $J$ \CSPS{}, and hence by Property \ref{theorem blance L CSPS}, balances the subspace spanned by these bifurcations of the $J$ contrasts, and therefore, any linear combination of these bifurcated contrasts.

\section{Illustrations}\label{sec: illustration}
The algorithm in Supplementary materials provides an illustrative routine for creating balance on \CSPS{} of $J$ contrasts and as well as for assessing balance for the contrasts and their linear combinations. We refer to this routine as ``the Algorithm''.

\subsection{An artificial example}\label{sec: artificial eg}

Table \ref{tab: balance eg} displays an artificial dataset with 24 units, three treatments and three covariates. We implement the Algorithm using two contrasts $\mb{\lambda}_1 = (1/2, 1/2, -1)$ and $\mb{\lambda}_2 = (1, -1, 0)$. We show the balance results for a third contrast $\mb{\lambda}_3 = \mb{\lambda}_1 - 1/2\mb{\lambda}_2 = (0, 1, -1)$, which is a linear combination of $\mb{\lambda}_1$ and $\mb{\lambda}_2$. We briefly discuss the balance results and an alternative scenario where only the bifurcation of $\mb{\lambda}_2$ is used for balancing.

\begin{table}[!htp]
\small
\centering
\caption{A simple artificial example showing that balance on $c_1, c_2$ balances $\mb{\lambda_3}$ }\label{tab: balance eg}
\begin{tabular}{cccccccccccccccc} \toprule
$(X_{i1}, X_{i2},X_{i3})$ & $W_i$&$D_{i1}$ & $ D_{i2} $ &   $c_1$ & $c_2$ &$D_{i3}$& \multicolumn{2}{c}{Balance on $c_1, c_2$} & \multicolumn{2}{c}{Covariate differences} \\  
 & & \multicolumn{5}{c}{}& Propensity score&Subclass labels &$D_{i3} = 1$ & $D_{i3} = -1$ \\  \hline
\multirow{6}{*}{(1,1,1)} & 1 & 1 & 1 & \multirow{6}{*}{$\frac{1}{3}$} & \multirow{6}{*}{$\frac{1}{2}$}&0&\multirow{6}{*}{$\frac{1}{5}$}&\multirow{6}{*}{$1$} &\multirow{6}{*}{$(1,1,1)$}&\multirow{6}{*}{$(1,1,1)$}\\
& 2  & 1  & -1 &                     &                     &1& \\
& 3  & -1 &0  &                     &                   &-1&  \\
& 3  & -1 & 0 &                     &                   &-1&  \\
& 3  & -1 & 0 &                     &                    &-1& \\
& 3  & -1 & 0  &                    &                    & -1&\\ \hline
\multirow{6}{*}{(1,0,1)} & 1 & 1 & 1 & \multirow{6}{*}{$\frac{2}{3}$} & \multirow{6}{*}{$\frac{3}{4}$}& 0&\multirow{6}{*}{$\frac{1}{3}$}& \multirow{6}{*}{$2$}&\multirow{6}{*}{$(1,0,1)$}&\multirow{6}{*}{$(1,0,1)$}\\
& 1 & 1  & 1  &               &&0&                   \\
& 1 & 1  & 1  &                   &&0&               \\
& 2 & 1  & -1 &                  & &1&                \\
& 3 & -1 & 0 &                    &&-1&                \\
& 3& -1 & 0  &                  &&-1&                    \\  \hline
\multirow{6}{*}{(0,1,1)} & 1 & 1 & 1 &\multirow{6}{*}{$\frac{5}{6}$} &\multirow{6}{*}{$\frac{2}{5}$}&0&\multirow{6}{*}{$\frac{3}{4}$}&\multirow{6}{*}{$3$} &\multirow{6}{*}{$(0,1,1)$}&\multirow{6}{*}{$(0,1,1)$}\\
& 1 & 1  & 1  &                    &&0                    \\
& 2 & 1  & -1 &                     & &1                  \\
& 2  & 1  & -1 &                     & &1                 \\
& 2 & 1  & -1 &                    &  &1                  \\
& 3  & -1 & 0  &                  & &-1                   \\  \hline
\multirow{6}{*}{(0,0,0)} & 1 &1  &1&\multirow{6}{*}{$\frac{2}{3}$} & \multirow{6}{*}{$\frac{1}{2}$}&0& \multirow{6}{*}{$\frac{1}{2}$}&\multirow{6}{*}{$4$}&\multirow{6}{*}{$(0,0,0)$}&\multirow{6}{*}{$(0,0,0)$}\\
& 1 & 1  & 1 &                   &&0                    \\
& 2 & 1  & -1 &                   & &1                  \\
& 2 & 1  & -1 &                    & &1                \\
& 3 & -1 & 0  &                &    &-1                \\ 
& 3 &-1 & 0  &                     &  &-1             \\ \bottomrule           
\end{tabular}
\end{table}

 The left panel of the table shows the treatment group indicators with respect to the two contrasts $\mb{\lambda}_1$ and $\mb{\lambda}_2$. After $c_1$ and $c_2$ are created, they are treated as covariates and a usual propensity score analysis is used to estimate the probability of $D_{i3} = 1$ versus $D_{i3} =-1$ given $(c_1(X_i),c_2(X_i))$. The right panel shows the estimated propensity scores and the subclass labels created by this algorithm. Clearly, within each subclass, the difference in means between treatment groups with respect to contrast $\mb{\lambda}_3$ is exactly balanced because balancing on $(c_1, c_2)$ balances the subspace spanned by these bifurcations of the two contrasts $\mb{\lambda}_1$ and $\mb{\lambda}_2$.

In contrast, balancing on the \CSPS{} of bifurcations does not balance the subspace that is orthogonal to these bifurcations. To illustrate this, we balance on $c_2$ and evaluate the balance performance with respect to contrast $\mb{\lambda}_1$. In the subclass with $c_2 = 1/2$, the covariate means for treatment groups with $D_{i1} = 1$ and $D_{i1} = -1$ are $(1/3,1/3,1/3)$ and $(2/3,2/3,2/3)$ respectively, which shows that the covariate difference between treatment groups is not $(0,0,0)$ at the same level of $c_2$.

\subsection{Simulation studies}\label{sec simulation}
In \S\ref{sec: artificial eg}, we were able to stratify the units into subclasses such that each subclass has only one level of $(c_1, c_2)$, but this is usually not the case when there are many values of $(c_1, c_2)$. We now consider such a case with $T=3$ and $N =  800$ units and covariates $X_{ik}\sim \textbf{Norm}(0,1)$, $k = 1, 2, 3$. The assignment mechanism is a multinomial logistic model, with $\Prob{W_i = t\mid X_i} = \exp(\beta_t'X_i)/\sum_{\ell = 1}^3 \exp(\beta_\ell'X_i)$; We consider two treatment assignment mechanisms: Mechanism I has $\mb{\beta}_1= \mb{\beta}_2 =\mb{\beta}_3 = (0,0,0)$, i.e., a complete randomization, and Mechanism II has $\mb{\beta}_1= (0,0,0),\mb{\beta}_2 = (0.75, 0.25, 0.5), \mb{\beta}_3 =(0.25, 0.75, 0.5)$. We consider four contrasts: $\mb{\lambda}_1 = (1/3, 2/3, -1)$, $ \mb{\lambda}_2 = (1, -1, 0)$, $\mb{\lambda}_3 = (1, 0, -1)$, $\mb{\lambda}_4=(0, 1,-1)$. We implement the Algorithm, balancing on the sgn bifurcation \CSPS{} of $\mb{\lambda}_1$ and $\mb{\lambda}_2$.

We repeat the simulations 100 times and report the results in Table \ref{tab: simulation0-1 CSPS}. For Experiment I, a completely randomized experiment, a small reduction is observed after balancing showing that the estimated \CSPS{} can reduce the random imbalance. We find that, for Experiment II, after implementing the Algorithm with simple subclassification, the differences in covariate means are substantially diminished, to less than 0.1 on average. 
\begin{table}[]
\small
\centering
\caption{ The mean difference in covariates before and after balancing}\label{tab: simulation0-1 CSPS}
\begin{tabular}{lrrrrrrrrrrrrr}
\toprule
Contrast & \multicolumn{6}{c}{Experiment I}                                       & \multicolumn{6}{c}{Experiment II}                                      \\ 
         & \multicolumn{3}{c}{Before balancing} & \multicolumn{3}{c}{After balancing} & \multicolumn{3}{c}{Before balancing} & \multicolumn{3}{c}{After balancing} \\
                  & $X_1$               & $X_2$  & $X_3$             & $X_1$              & $X_2$     & $X_3$       & $X_1$             & $X_2$     & $X_3$         & $X_1$             & $X_2$     & $X_3$       \\ \midrule
$\lambda_1$  & 0.00           & 0.01  & 0.02 & 0.00          & 0.00 & 0.00 & 0.11           & -0.79 & -0.21 & 0.04          & -0.08 & 0.00  \\
$\lambda_2$  & -0.01          & -0.01 & 0.00 & 0.00          & 0.00 & 0.00 & -0.46          & -0.23 & -0.46 & -0.06         & 0.01  & -0.05 \\
$\lambda_3$  & -0.01          & -0.01 & 0.00 & 0.00          & 0.00 & 0.00 & -0.46          & -0.23 & -0.46 & -0.06         & 0.01  & -0.05 \\
$\lambda_4$  & 0.01           & 0.01  & 0.01 & 0.00          & 0.00 & 0.00 & 0.57           & -0.54 & 0.24  & 0.08          & -0.07 & 0.03           \\ \bottomrule
\end{tabular}
\end{table}

\section{Discussion}\label{sec: discussion}
\CSPS{} methodology focuses on creating treatment groups with balanced covariates in design. Once balanced groups are created, treatment effects can be robustly estimated. For example, recent work suggests that weighting estimators are generally worse than imputation based estimators \citep[e.g.][]{GR13,GR17}, advice which can be traced back to \cite{Rubin73}.

\section*{Acknowledgement}
We thank the referees for helpful comments and Rui Dong and Ke Zhu for helpful discussions. 

\section*{Supplementary material}
\label{SM}
Supplementary material includes the proofs for Property 1 and 2, the details of the Algorithm.\bibliographystyle{biometrika}
\bibliography{cspsbmref/PropensityScorePersonalized2,cspsbmref/csps2,cspsbmref/CausalSeq}

\begin{thebibliography}{16}
\expandafter\ifx\csname natexlab\endcsname\relax\def\natexlab#1{#1}\fi

\bibitem[{Gutman \& Rubin(2017)}]{GR17}
\textsc{Gutman, R.} \& \textsc{Rubin, D.} (2017).
\newblock Estimation of causal effects of binary treatments in unconfounded
  studies with one continuous covariate.
\newblock \textit{Statistical methods in medical research} \textbf{26},
  1199--1215.

\bibitem[{Gutman \& Rubin(2013)}]{GR13}
\textsc{Gutman, R.} \& \textsc{Rubin, D.~B.} (2013).
\newblock Robust estimation of causal effects of binary treatments in
  unconfounded studies with dichotomous outcomes.
\newblock \textit{Statistics in medicine} \textbf{32}, 1795--1814.

\bibitem[{Horvitz \& Thompson(1952)}]{HT52}
\textsc{Horvitz, D.~G.} \& \textsc{Thompson, D.~J.} (1952).
\newblock A generalization of sampling without replacement from a finite
  universe.
\newblock \textit{Journal of the American statistical Association} \textbf{47},
  663--685.

\bibitem[{Imai \& van Dyk(2004)}]{ID04}
\textsc{Imai, K.} \& \textsc{van Dyk, D.~A.} (2004).
\newblock Causal inference with general treatment regimes.
\newblock \textit{Journal of the American Statistical Association} \textbf{99},
  854--866.

\bibitem[{Imbens(2000)}]{Imbens00}
\textsc{Imbens, G.~W.} (2000).
\newblock The role of the propensity score in estimating dose-response
  functions.
\newblock \textit{Biometrika} \textbf{87}, 706--710.

\bibitem[{Imbens \& Rubin(2015)}]{IR15}
\textsc{Imbens, G.~W.} \& \textsc{Rubin, D.~B.} (2015).
\newblock \textit{Causal inference in statistics, social, and biomedical
  sciences}.
\newblock Cambridge University Press.

\bibitem[{Kaplan et~al.(2019)Kaplan, Mashash, Williams, Batchelder, Starr-Glass
  \& Zeitzer}]{KMW+19}
\textsc{Kaplan, K.}, \textsc{Mashash, M.}, \textsc{Williams, R.},
  \textsc{Batchelder, H.}, \textsc{Starr-Glass, L.} \& \textsc{Zeitzer, J.}
  (2019).
\newblock Effect of light flashes vs sham therapy during sleep with adjunct
  cognitive behavioral therapy on sleep quality among adolescents: a randomized
  clinical trial.
\newblock \textit{JAMA Netw Open} \textbf{2}, e1911944.

\bibitem[{LaLonde(1986)}]{Lalonde86}
\textsc{LaLonde, R.~J.} (1986).
\newblock Evaluating the econometric evaluations of training programs with
  experimental data.
\newblock \textit{The American economic review} , 604--620.

\bibitem[{Lechner(2001)}]{Lechner01}
\textsc{Lechner, M.} (2001).
\newblock Identification and estimation of causal effects of multiple
  treatments under the conditional independence assumption.
\newblock \textit{In Econometric evaluation of labour market policies, Ed. M.
  Lechner and F. Pfeiffer. Physica, Heidelberg} , 43--58.

\bibitem[{Rosenbaum \& Rubin(1983)}]{RR83}
\textsc{Rosenbaum, P.~R.} \& \textsc{Rubin, D.~B.} (1983).
\newblock The central role of the propensity score in observational studies for
  causal effects.
\newblock \textit{Biometrika} \textbf{70}, 41--55.

\bibitem[{Rosenthal et~al.(2000)Rosenthal, Rosnow \& Rubin}]{RRR00}
\textsc{Rosenthal, R.}, \textsc{Rosnow, R.~L.} \& \textsc{Rubin, D.~B.} (2000).
\newblock \textit{Contrasts and effect sizes in behavioral research: A
  correlational approach}.
\newblock Cambridge University Press.

\bibitem[{Rubin(1974)}]{Rubin74}
\textsc{Rubin, D.} (1974).
\newblock Estimating causal effects of treatments in randomized and
  nonrandomized studies.
\newblock \textit{Journal of Educational Psychology} \textbf{66}, 688.

\bibitem[{Rubin(1973)}]{Rubin73}
\textsc{Rubin, D.~B.} (1973).
\newblock Matching to remove bias in observational studies.
\newblock \textit{Biometrics} , 159--183.

\bibitem[{Rubin(1980)}]{Rubin80b}
\textsc{Rubin, D.~B.} (1980).
\newblock Discussion of "randomization analysis of experimental data in the
  fisher randomization test.
\newblock \textit{Journal of the American Statistical Association} \textbf{75},
  591--593.

\bibitem[{Snedecor \& Cochran(1967)}]{SC67}
\textsc{Snedecor, G.~W.} \& \textsc{Cochran, W.~G.} (1967).
\newblock \textit{Statistical Methods}.
\newblock Iowa State Univerisity Press, 1st Edition.

\bibitem[{Wu \& Hamada(2011)}]{WH11}
\textsc{Wu, C.~J.} \& \textsc{Hamada, M.~S.} (2011).
\newblock \textit{Experiments: planning, analysis, and optimization}, vol. 552.
\newblock John Wiley \& Sons.

\end{thebibliography}
\end{document}